\documentstyle[aps,twocolumn,prl,amssymb]{revtex}
\tighten
\begin{document}
\twocolumn[\hsize\textwidth\columnwidth\hsize\csname @twocolumnfalse\endcsname
\draft
\title{Five Quantum Register Error Correction Code For Higher Spin Systems}
\author{H. F. Chau\footnote{e-mail: hfchau@hkusua.hku.hk}}
\address{Department of Physics, University of Hong Kong, Pokfulam Road, Hong
Kong}
\date{\today}
\preprint{HKUPHYS-HFC-02; quant-ph/9702033}
\maketitle
\begin{abstract}
 I construct a quantum error correction code (QECC) in higher spin systems
 using the idea of multiplicative group character. Each $N$ state quantum
 particle is encoded as five $N$ state quantum registers. By doing so, this
 code can correct any quantum error arising from any one of the five quantum
 registers. This code generalizes the well-known five qubit perfect code in
 spin-1/2 systems and is shown to be optimal for higher spin systems. I also
 report a simple algorithm for encoding. The importance of multiplicative group
 character in constructing QECCs will be addressed.
\end{abstract}
\medskip
]
\narrowtext
\footnotetext[1]{e-mail: hfchau@hkusua.hku.hk}
 The power of a quantum computer is perhaps best illustrated by the powerful
 Shor's quantum polynomial time factorization algorithm \cite{Shor94}. However,
 the real power of a quantum computer may be much more limited because it is
 extremely vulnerable to disturbance \cite{Landauer94}. Nevertheless, Shor
 pointed out later that the effect of quantum decoherence can be compensated if
 we introduce redundancy in the quantum state in a suitable way. We first
 encode the quantum state into a larger Hilbert space $H$. Then we measure the
 wave function in a suitable subspace $C$ of $H$. And finally we apply a
 unitary transformation to the orthogonal complement of $C$ according to our
 measurement result, it is possible to correct quantum errors due to
 decoherence with the environment \cite{Shor95}. This kind of scheme is now
 called quantum error correction code (QECC).
\par
 Since then, many QECCs have been discovered (see, for example,
 Refs.~\cite{Laflamme96,CS96,St96a,Steane96,Calderbank96,Gottesman96,Chau97})
 and various theories on QECC have also been developed (see, for example,
 Refs.~\cite{Steane96,Calderbank96,Gottesman96,Chau97,Steane96a,Knill96a,Knill96,Bennett96}).
 In particular, the necessary and sufficient condition for a QECC is
 \cite{Knill96a,Knill96,Bennett96}
\begin{equation}
 \langle i_{\rm Encode} | A^{\dag} \! B |j_{\rm Encode} \rangle = \lambda_{A,B}
 \delta_{ij} ~, \label{E:Condition}
\end{equation}
 where $|i_{\rm Encode} \rangle$ denotes the encoded quantum state $|i\rangle$
 using the QECC, $A, B$ are the possible errors that can be handled by the
 QECC, and $\lambda_{A,B}$ is a complex constant independent of
 $|i_{\rm Encode}\rangle$ and $|j_{\rm Encode}\rangle$.
\par
 Early QECCs deal with decoherence of individual spin-1/2 particles with the
 environment. Besides, the information loss to the environment is assumed to be
 unrecoverable. More recently, Duan and Guo considered the decoherence of
 spin-1/2 particles with the same environment. Based on a specific model of the
 environment in thermal equilibrium, they found a new coding scheme
 \cite{Duan_Guo}. Another investigation concentrates on the mutual decoherence
 between the quantum spins inside the quantum computer. Chau pointed out that
 the ability to correct quantum errors between various registers inside a
 quantum computer is equivalent to the ability to correct quantum error of a
 single quantum higher spin particle \cite{Chau97}. Thus, it is interesting to
 construct QECCs for quantum registers with higher spin.
\par
 The first QECC for particles with spin higher than 1/2 was found by Chau using
 group-theoretical methods. He encodes each quantum particle as nine quantum
 registers. And by doing so, his code can correct any quantum error involving
 exactly one quantum register \cite{Chau97}. Nonetheless, his code is not
 perfect.\footnote{See Ref.~\cite{Laflamme96} for a precise definition of a
 perfect code.} So, it is nature to ask if it is possible to construct more
 economical codes for higher spin systems.
\par
 An affirmative answer is provided in this paper. I report a way to encode
 each quantum particle as five quantum registers, which can correct error in
 at most one of the five registers. I also show that this code is optimal in
 the sense that no QECC with codeword length less than five can correct
 general one quantum register error. For spin-1/2 particles, this code is
 equivalent to the perfect codes discovered by Laflamme {\em et al.}
 \cite{Laflamme96} and Bennett {\em et al.} \cite{Bennett96} up to unitary
 transformations. As you will see in the derivation, the success of this five
 register code relies heavily on the sum rule of the multiplicative group
 character of the finite additive group ${\Bbb Z}_N$.
\par
 The (multiplicative) group character of the finite additive group ${\Bbb Z}_N$
 is a map $\chi : {\Bbb Z}_N \longrightarrow {\Bbb C}$ satisfying \cite{NT}
\begin{equation}
 \chi (a + b) = \chi (a) \chi (b) \label{E:Group_Character}
\end{equation}
 for all $a,b\in {\Bbb Z}_N$. Then $\chi$ satisfies the sum rule \cite{NT}
\begin{equation}
 \sum_{m\in {\Bbb Z}_N} \chi (m) = \left\{ \begin{array}{ll} N &
 \hspace{0.15in} \mbox{if~} \chi \mbox{~is~the~trivial~character,} \\ 0 &
 \hspace{0.15in} \mbox{otherwise.} \end{array} \right. \label{E:Sum_Rule}
\end{equation}
 More concretely, the above sum rule can be written as
\begin{equation} 
 \sum_{m=0}^{N-1} \omega_N^{mk} = \left\{ \begin{array}{ll} N & \hspace{0.1in}
 \mbox{if~} k = 0 \mbox{~mod~} N, \\ 0 & \hspace{0.1in} \mbox{for~} k = 1,2,
 \ldots ,N-1 \mbox{~mod~} N, \end{array} \right. \label{E:omega}
\end{equation}
 where $\omega_N$ is a primitive $N$-th root of unity.
\par
 To see how we use Eq.~(\ref{E:omega}) to construct our five quantum register
 code, let us begin by denoting the $N$ mutually orthogonal eigenstates in each
 quantum register by $0\rangle$, $|1\rangle$, \ldots , $|N-1\rangle$. Then, I
 claim that the following encoding scheme can correct any quantum error
 occurring in at most one of the quantum registers
\begin{eqnarray}
 |k\rangle & \longmapsto & \frac{1}{N^{3/2}} \!\!\sum_{p,q,r = 0}^{N-1}
 \!\!\omega_N^{k(p+q+r) + p r} |p\!+\!q\!+\!k\rangle \otimes |p\!+\!r\rangle
 \otimes \nonumber \\ & & ~~~~ |q\!+\!r\rangle \otimes |p\rangle \otimes
 |q\rangle \nonumber \\ & \equiv & \frac{1}{N^{3/2}} \!\!\sum_{p,q,r = 0}^{N-1}
 \!\!\omega_N^{k(p+q+r) + p r} |p\!+\!q\!+\!k, p\!+\!r, q\!+\!r, p, q\rangle
 \label{E:Encoding}
\end{eqnarray}
 for $k = 0,1,\ldots ,N-1$, where all the additions in the state kets and in
 the sum are modulo $N$.
\par
 Let me denote the one bit quantum error $E_\alpha$ occurring at the $i$-th
 quantum register by the symbol $E_{i,\alpha}$. To prove the above claim, it
 suffices to show that Eq.~(\ref{E:Condition}) holds for any quantum errors $A
 = E_{i,\alpha}$ and $B = E_{j,\beta}$ for $1 \leq i \leq j \leq 5$.
\par
 First, I consider the case when $(i,j)$ = $(1,4)$ as a warm up. We have
\begin{eqnarray}
 & & \langle k_{\rm Encode} | E_{1,\alpha}^{\dag} E_{4,\beta} | k'_{\rm Encode}
 \rangle \nonumber \\ & = & \frac{1}{N^3} \!\!\!\sum_{p,q,r,p',q',r' = 0}^{N-1}
 \!\!\!\omega_N^{k'(p'+q'+r') + p' r' - k(p+q+r) - p r} \times \nonumber \\ & &
 ~~~ \langle p\!+\!q\!+\!k | E_\alpha^{\dag} | p'\!+\!q'\!+\!k' \rangle
 \,\langle p\!+\!r | p'\!+\!r' \rangle \,\langle q\!+\!r | q'\!+\!r' \rangle
 \times \nonumber \\ & & ~~~ \langle p | E_\beta | p' \rangle \,\langle q | q'
 \rangle \nonumber \\ & = & \frac{1}{N^3}
 \!\!\!\!\sum_{p,q,r,p',q',r' = 0}^{N-1} \!\!\!\!\omega_N^{k'(p'+q'+r') + p' r'
 - k(p+q+r) - p r} \,\delta_{q+r,q'+r'} \nonumber \\ & & ~~~ \delta_{q,q'}
 \,\delta_{p+r,p'+r'} \,\langle p\!+\!q\!+\!k | E_\alpha^{\dag} |
 p'\!+\!q'\!+\!k' \rangle \,\langle p | E_\beta | p' \rangle \nonumber \\ & = &
 \frac{1}{N^3} \!\!\!\sum_{p,q,r = 0}^{N-1} \!\!\!\omega_N^{(k' - k)(p+q+r)}
 \langle p\!+\!q\!+\!k | E_\alpha^{\dag} | p\!+\!q\!+\!k' \rangle \langle p |
 E_\beta | p \rangle \nonumber \\ & = & \delta_{k,k'} \,\frac{1}{N^2}
 \!\sum_{p,q = 0}^{N-1} \langle p\!+\!q | E_\alpha^{\dag} | p\!+\!q \rangle
 \,\langle p | E_\beta | p \rangle \nonumber \\ & \equiv & \delta_{k,k'}
 \,\Lambda_{1,\alpha;4,\beta} ~, \label{E:k_14}
\end{eqnarray}
 where $\Lambda_{1,\alpha;4,\beta}$ is independent of $k$. Thus,
 Eq.~(\ref{E:Condition}) hold when $(i,j)$ = $(1,4)$. Using the same trick, it
 is easy to verify that Eq.~(\ref{E:Condition}) holds when $(i,j)$ = $(1,1)$,
 $(2,2)$, $(3,3)$, $(4,4)$, $(5,5)$, $(1,5)$, and $(3,5)$.
\par
 Now, I proceed to the more difficult case when $(i,j)$ = $(1,2)$. We have
\begin{eqnarray}
 & & \langle k_{\rm Encode} | E_{1,\alpha}^{\dag} E_{2,\beta} | k'_{\rm Encode}
 \rangle \nonumber \\ & = & \frac{1}{N^3} \!\!\!\sum_{p,q,r,p',q',r' = 0}^{N-1}
 \!\!\!\omega_N^{k'(p'+q'+r') + p' r' - k(p+q+r) - p r} \delta_{p,p'}
 \,\delta_{q,q'} \nonumber \\ & & ~~~ \delta_{q+r,q'+r'} \langle p\!+\!q\!+\!k
 | E_\alpha^{\dag} | p'\!+\!q'\!+\!k' \rangle \,\langle p\!+\!r | E_{\beta} |
 p'\!+\!r' \rangle \nonumber \\ & = & \frac{1}{N^3} \!\sum_{p,q,r = 0}^{N-1}
 \!\omega_N^{(k' - k)(p+q+r)} \langle p\!+\!q\!+\!k | E_\alpha^{\dag} |
 p\!+\!q\!+\!k' \rangle \times \nonumber \\ & & ~~~ \langle p\!+\!r | E_\beta |
 p\!+\!r \rangle ~.
 \label{E:k_12_a}
\end{eqnarray}
 By relabeling $x = p + q$, $y = p + r$, and $z = r$, Eq.~(\ref{E:k_12_a}) can
 be rewritten as
\begin{eqnarray}
 & & \langle k_{\rm Encode} | E_{1,\alpha}^{\dag} E_{2,\beta} | k'_{\rm Encode}
 \rangle \nonumber \\ & = & \frac{1}{N^3} \!\sum_{x,y,z = 0}^{N-1}
 \!\omega_N^{(k' - k)(x + z)} \langle x\!+\!k | E_\alpha^{\dag} | x\!+\!k'
 \rangle \,\langle y | E_\beta | y \rangle \nonumber \\ & = & \delta_{k,k'}
 \,\frac{1}{N^2} \!\sum_{x,y = 0}^{N-1} \langle x | E_\alpha^{\dag} | x \rangle
 \,\langle y | E_\beta | y \rangle \nonumber \\ & \equiv & \delta_{k,k'}
 \,\Lambda_{1,\alpha;2,\beta} ~, \label{E:k_12}
\end{eqnarray}
 where $\Lambda_{1,\alpha;2,\beta}$ is independent of $k$. Thus,
 Eq.~(\ref{E:Condition}) holds when $(i,j)$ = $(1,2)$. In a similar way, one
 can show that Eq.~(\ref{E:Condition}) is also true for $(i,j)$ = $(1,3)$.
\par
 Now, I move on to the case when $(i,j)$ = $(2,3)$. By direct computation, we
 obtain
\begin{eqnarray}
 & & \langle k_{\rm Encode} | E_{2,\alpha}^{\dag} E_{3,\beta} | k'_{\rm Encode}
 \rangle \nonumber \\ & = & \frac{1}{N^3} \!\!\!\sum_{p,q,r,p',q',r' = 0}^{N-1}
 \!\!\!\omega_N^{k'(p'+q'+r') + p' r' - k(p+q+r) - p r} \delta_{p,p'}
 \,\delta_{q,q'} \nonumber \\ & & ~~~ \delta_{p+q+k,p'+q'+k'} \langle p\!+\!r |
 E_\alpha^{\dag} | p'\!+\!r' \rangle \,\langle q\!+\!r | E_\beta | q'\!+\!r'
 \rangle \nonumber \\ & = & \delta_{k,k'} \,\frac{1}{N^3}
 \!\!\!\sum_{p,q,r,r' = 0}^{N-1} \!\!\!\omega_N^{(r' - r)(k+p)} \langle
 p\!+\!r | E_\alpha^{\dag} | p\!+\!r' \rangle \times \nonumber \\ & & ~~~
 \langle q\!+\!r | E_\beta | q\!+\!r' \rangle ~. \label{E:k_23_a}
\end{eqnarray}
 By relabeling $x = r' - r$, $y = p+r$, $z = q+r$, and $u = p$,
 Eq.~(\ref{E:k_23_a}) can be written as
\begin{eqnarray}
 & & \langle k_{\rm Encode} | E_{2,\alpha}^{\dag} E_{3,\beta} | k'_{\rm Encode}
 \rangle \nonumber \\ & = & \delta_{k,k'} \,\frac{1}{N^3}
 \!\!\sum_{u,x,y,z = 0}^{N-1} \!\!\omega_N^{x(u+k)} \langle y | E_\alpha^{\dag}
 | y\!+\!x \rangle \,\langle z | E_\beta | z\!+\!x \rangle \nonumber \\ & = &
 \delta_{k,k'} \,\frac{1}{N^2} \!\sum_{x,y,z = 0}^{N-1} \langle y |
 E_\alpha^{\dag} | y\!+\!x \rangle \,\langle z | E_\beta | z\!+\!x \rangle
 \nonumber \\ & \equiv & \delta_{k,k'} \,\Lambda_{2,\alpha;3,\beta} ~,
 \label{E:k_23}
\end{eqnarray}
 where $\Lambda_{2,\alpha;3,\beta}$ is independent of $k$. Hence,
 Eq.~(\ref{E:Condition}) is also satisfied when $(i,j)$ = $(2,3)$. Using
 similar methods, it can be shown that Eq.~(\ref{E:Condition}) holds if $(i,j)$
 = $(2,4)$, $(2,5)$, and $(3,4)$.
\par
 Finally, I consider the case when $(i,j)$ = $(4,5)$. By direct computation,
 we find that
\begin{eqnarray}
 & & \langle k_{\rm Encode} | E_{4,\alpha}^{\dag} E_{5,\beta} | k'_{\rm Encode}
 \rangle \nonumber \\ & = & \frac{1}{N^3} \!\!\!\sum_{p,q,r,p',q',r' = 0}^{N-1}
 \!\!\!\omega_N^{k'(p'+q'+r') + p' r' - k(p+q+r) - p r} \delta_{p+r,p'+r'}
 \nonumber \\ & & ~~~~\delta_{q+r,q'+r'} \delta_{p+q+k,p'+q'+k'} \,\langle p |
 E_\alpha^{\dag} | p' \rangle \,\langle q | E_\beta | q' \rangle \nonumber \\ &
 = & \frac{1}{N^3} \!\!\!\!\!\sum_{p,q,r,p',q',r' = 0}^{N-1}
 \!\!\!\!\!\omega_N^{k'(p' + q' + r') + p' r' - k(p+q+r) - p r}
 \delta_{2p+k,2p'+k'} \nonumber \\ & & ~~~ \delta_{2q+k,2q'+k'}
 \,\delta_{2r-k,2r'-k'} \,\langle p | E_\alpha^{\dag} | p' \rangle \,\langle q
 | E_\beta | q' \rangle ~. \label{E:k_45_a}
\end{eqnarray}
 Let us analyze the situation by considering the following two sub-cases:
\par\noindent
{\em Sub-case (a):} If $k - k'$ is odd and $N$ is even, then it is impossible
 to find $p, p' \in {\Bbb Z}_N$ such that $2p + k = 2p' + k' \mbox{~mod~} N$.
 Hence, the existence of the $\delta_{2p+k,2p'+k'}$ term in
 Eq.~(\ref{E:k_45_a}) implies that $\langle k_{\rm Encode} |
 E_{4,\alpha}^{\dag} E_{5,\beta} | k'_{\rm Encode} \rangle = 0$.
\par\noindent
{\em Sub-case (b):} if either $k - k'$ is even or $N$ is odd, then it is
 possible to find $p, p' \in {\Bbb Z}_N$ such that $2p + k= 2p' + k'
 \mbox{~mod~} N$. That is to say, it make sense to regard $(k' - k) / 2$ as an
 integer in ${\Bbb Z}_N$. Then Eq.~(\ref{E:k_45_a}) becomes
\begin{eqnarray}
 & & \langle k_{\rm Encode} | E_{4,\alpha}^{\dag} E_{5,\beta} | k'_{\rm Encode}
 \rangle \nonumber \\ & = & \frac{1}{N^3} \!\!\sum_{p,q,r = 0}^{N-1}
 \!\!\omega_N^{\left[ \frac{k' - k}{2} \right] (3p + 2q + r - \frac{3k' - k}{2}
 )} \langle p | E_\alpha^{\dag} | p - \frac{k' - k}{2} \rangle \times \nonumber
 \\ & & ~~~ \langle q | E_\beta | q - \frac{k' - k}{2} \rangle \nonumber \\ & =
 & \delta_{k,k'} \,\frac{1}{N^2} \!\sum_{p,q = 0}^{N-1} \langle p |
 E_\alpha^{\dag} | p \rangle \,\langle q | E_\beta | q \rangle \nonumber \\ &
 \equiv & \delta_{k,k'} \,\Lambda_{4,\alpha;5,\beta} ~, \label{E:k_45}
\end{eqnarray}
 where $\Lambda_{4,\alpha;5,\beta}$ is independent of $k$.
\par
 Therefore, the encoding scheme in Eq.~(\ref{E:Encoding}) satisfies
 Eq.~(\ref{E:Condition}) for any $(i,j)$ with $1 \leq i,j \leq 5$. And hence,
 this scheme is able to correct any quantum error arising at any one of the
 quantum registers as promised.
\par
 The key idea used in this five register code is (i) the multiplicative group
 character sum rule in Eq.~(\ref{E:Sum_Rule}), (ii) the relabeling of some
 variables in the summation, and (iii) the strong correlation between the five
 quantum registers.\footnote{In order words, the high entanglement entropy in
 this code.} Since the sum rule in Eq.~(\ref{E:Sum_Rule}) plays a very
 important role in both the five and the nine quantum register codes
 \cite{Chau97}, it will be interesting to rewrite other existing QECCs for
 spin-1/2 particles in a form similar to that of Eq.~(\ref{E:Encoding}). This
 may provide a way to generalize these codes to higher spin systems.
\par
 Back to the case when $N = 2$. The above encoding scheme above can be
 explicitly written as
\begin{mathletters}
\begin{eqnarray}
 |0\rangle & \longmapsto & \frac{1}{\sqrt{8}} \left[ |00000\rangle +
 |01100\rangle + |10101\rangle + |11001\rangle \right. \nonumber \\ & & ~~~
 \left. + |11010\rangle - |10110\rangle + |01111\rangle - |00011\rangle \right]
 ~, \label{E:Encoding_N2_0}
\end{eqnarray}
and
\begin{eqnarray}
 |1\rangle & \longmapsto & \frac{1}{\sqrt{8}} \left[ |10000\rangle -
 |11100\rangle - |00101\rangle + |01001\rangle \right. \nonumber \\ & & ~~~
 \left. - |01010\rangle - |00110\rangle + |11111\rangle + |10011\rangle \right]
 ~. \label{E:Encoding_N2_1}
\end{eqnarray}
\end{mathletters}
\par\noindent
This scheme can be transformed to the perfect code obtained by Laflamme
 {\em et al.} \cite{Laflamme96} (and hence also Bennett {\em et al.}'s
 \cite{Bennett96}) by a simple unitary transformation: first permute the
 five quantum registers by $P(13524)$, then adds an extra phase of $\pi$ to the
 encoding state whenever $p+r+k$ is even. That is to say, Laflamme
 {\em et al.}'s perfect code can be written as
\begin{eqnarray}
 |k\rangle & \longmapsto & \frac{1}{\sqrt{8}} \!\sum_{p,q,r}
 (-1)^{(p+1)(r+1)+k(p+q+r+1)} |p\!+\!q\!+\!1\rangle \otimes \nonumber \\ & &
 ~~~~ |p\rangle \otimes |p\!+\!r\rangle \otimes |q\rangle \otimes |q\!+\!r
 \rangle ~, \label{E:Laflamme_Perfect_Code}
\end{eqnarray}
 for $k = 0, 1$.
\par
 Now, I give a simple encoding algorithm for this code. Using a series of
 quantum binary conditional-NOT gates, we may ``copy'' the state
 $|k,0,0,0,0\rangle$ to $|k,0,k,k,k\rangle$ efficiently. Then, we apply
 quantum discrete Fourier transforms similar to that used in Shor's
 algorithm \cite{Shor94,Ekert96,Hoyer97} separately to the third, fourth, and
 fifth quantum registers. Then, we add an additional phase of $\omega_N^{p r}$
 to the system using a Toffoli-like gate \cite{Chau95,Barenco95}. We then use
 a series of quantum binary conditional-NOT gates to ``copy'' the fourth
 register to the second one. Finally, by suitably adding the quantum registers
 together using reversible quantum logic gates, we obtain the quantum code as
 required. The entire encoding procedure can be summarized below:
\begin{eqnarray}
 & & |k,0,0,0,0\rangle \nonumber \\ & \longmapsto & |k,0,k,k,k\rangle \nonumber
 \\ & \longmapsto & \frac{1}{N^{3/2}} \!\sum_{p,q,r = 0}^{N-1}
 \!\!\omega_N^{k(p+q+r)} |k,0,r,p,q\rangle \nonumber \\ & \longmapsto &
 \frac{1}{N^{3/2}} \!\sum_{p,q,r = 0}^{N-1} \!\!\omega_N^{k(p+q+r) + p r}
 |k,0,p,q,r\rangle \nonumber \\ & \longmapsto & \frac{1}{N^{3/2}}
 \!\sum_{p,q,r = 0}^{N-1} \!\!\omega_N^{k(p+q+r) + p r} |k,p,r,p,q\rangle
 \nonumber \\ & \longmapsto & \frac{1}{N^{3/2}} \!\sum_{p,q,r = 0}^{N-1}
 \!\!\omega_N^{k(p+q+r) + p r } |p\!+\!q\!+\!k,p\!+\!r,q\!+\!r,p,q\rangle ~.
 \label{E:Construction}
\end{eqnarray}
\par\indent
 Finally, I present a proof of the optimality of the above QECC. More
 precisely, I will show that it is not possible to correct a general quantum
 error involving exactly one quantum register by encoding each quantum particle
 by four (or less) quantum registers. Following Section~V-B in
 Ref.~\cite{Knill96a} (see also ref.~\cite{Grassl}), I suppose a single error
 correcting quantum code with codeword length four exists. Then one can always
 write
\begin{equation}
 |i_{\rm Encode}\rangle = \sum_{p,q,r,s = 0}^{N-1} \alpha^{(i)}_{pqrs}
 |p,q,r,s\rangle \label{E:Four_Bit}
\end{equation}
 for $i = 0,1,\ldots ,N-1$. Define the reduced density matrices
\begin{equation}
 \rho^{(i)}_{p'q';pq} = \sum_{r,s = 0}^{N-1} \left( \alpha^{(i)}_{p'q'rs}
 \right)^{*} \alpha^{(i)}_{pqrs} \label{E:Den_Reduced}
\end{equation}
 for all $i$, and
\begin{equation}
 \left( E_{(i_0,j_0)} \right)_{ij} = \left\{ \begin{array}{ll} 1 &
 \hspace{0.2in} \mbox{if~} i = i_0 \mbox{~and~} j = j_0, \\ 0 & \hspace{0.2in}
 \mbox{otherwise.} \end{array} \right. \label{E:Error_Op}
\end{equation}
 Now we consider the error operators $E_{3,(i_0,j_0)}$ and $E_{4,(i_0,j_0)}$
 which act on the third and forth register, respectively. Suppose
 $|i_{\rm Encode}\rangle \neq |j_{\rm Encode}\rangle$, then from
 Eqs.~(\ref{E:Condition}) and~(\ref{E:Four_Bit})--(\ref{E:Den_Reduced}), one
 arrives at
\begin{equation}
 \rho^{(i)} \rho^{(j)} = 0 \label{E:Sum_1}
\end{equation}
 for all $i\neq j$. Similarly, we consider the actions of $E_{1,(i_0,j_0)}$ and
 $E_{2,(i_0,j_0)}$ on the encoded registers. Putting $i = j$ in
 Eq.~(\ref{E:Condition}), one arrives at
\begin{equation}
 \rho^{(i)} = \rho^{(j)} \label{E:Same_1}
\end{equation} 
 for all $i, j$. From Eqs.~(\ref{E:Sum_1})--(\ref{E:Same_1}), one concludes
 that all the (hermitian) reduced density matrices $\rho^{(i)}$ are nilpotent.
 However, this is possible only if $\rho^{(i)} = 0$ and hence
 $\alpha^{(i)}_{pqrs} = 0$ for all $i,p,q,r,s = 0,1,\ldots ,N-1$. This
 contradicts with the assumption that $\sum_{p,q,r,s=0}^{N-1}
 \alpha^{(i)}_{pqrs} |p,q,r,s\rangle$ encodes the quantum state $|i\rangle$.
 Thus, the codeword length must be at least five. Consequently, the five
 quantum register code reported here is optimal.
\acknowledgements
 I would like to thank E. Knill and R. Laflamme for pointing out a mistake in
 the earlier version of this manuscript, M. Grassl for pointing out the use of
 Ref.\cite{Knill96a} to prove the optimality of this code. In addition,
 valuable discussions with M. Ben-Or is gratefully acknowledged.

\end{document}